\documentclass[12pt]{article}
\usepackage{amssymb}
\usepackage{amsfonts}
\usepackage{amsmath}
\usepackage{graphicx}

\newcommand{\be}{\begin{equation}}
\newcommand{\ee}{\end{equation}}
\newcommand{\ben}{\begin{eqnarray}}
\newcommand{\een}{\end{eqnarray}}

\newcommand{\nd}{{\noindent}}

\begin{document}
\begin{center}
{\Large Dynamics of the intensity-dependent Jaynes-Cummings model analyzed
via Fisher information}

S. Abdel-Khalek$^{a}$\footnote{%
E-mail: \textbf{sayedquantum@yahoo.co.uk,}}

$^{a}${\small Department of Mathematics, Faculty of Science, Sohag
University, 82524 Sohag, Egypt}\bigskip

A. Plastino$^{b}$\footnote{%
E-mail: \textbf{plastino@fisica.unlp.edu.ar}}

$^{b}${\small IFLP-CCT-Conicet, National University La Plata, 1900 La Plata,
Argentina }

A-S F. Obada$^{c}$

$^{c}$Department of Mathematics, Faculty of Science,Al-Azhar University,
Cairo, Egypt
\end{center}

\begin{quote}
{\noindent \textbf{Abstract}: } The dynamics of the Buck and
Sukumar model [B. Buck and C.V. Sukumar, Phys. Lett. A 81 (1981)
132] are investigated using different semi-classical
information-theory tools. Interesting aspects of the periodicity
inherent to the model are revealed and somewhat unexpected
features are revealed that seem to be related to the
classical-quantum transition.

\end{quote}


\section{Introduction}

{\noindent} The generation of nonclassical light and its
interaction with matter are receiving intense attention in quantum
optics, driven by research opportunities emerging from  present
control-technology regarding atoms and electromagnetic fields.
 The concomitant, all important details of the matter-field interaction have been thoroughly
scrutinized, with the Jaynes-Cummings model (JCM) playing a
pivotal role \cite{jc1,jc2,jc3,jc4}. Despite of the
JCM-simplicity, it permits a variety of generalizations,
applicable to distinct environments and regimes \cite{jc2}. In
particular, we may mention the work of Buck and Sukumar \cite{jc3}
that introduced the intensity dependent JCM. Because of the
commensurability of the Rabi frequencies arising from the model's
couplings, periodic revivals emerge, absent in the original JCM,
with a time-dependent state-vector that is periodic itself. As a
consequence, any expectation value will share such feature, which
leads to an enhancement of
certain effects that would otherwise be ignored by JCM-practitioners \cite%
{jc4,FVR98}. We wish here revisit the Buck-Sukumar model with
information-theory tools so as to be in a position to display
hopefully interesting details of the model's dynamics.

{\noindent} These tools have also been the subject of much
interest, in particular when they are applied in non-thermal
setting. In this regard, von Neumann's (NE) \cite{neum}, linear
(LE) \cite{manfredi}, and Shannon's entropy (SE) \cite{shann} have
been frequently used for a variety of quantum systems. It is worth
mentioning that the SE involves only the diagonal elements of the
density matrix and in some cases yields information similar to
that obtained from either the NE or LE measures. Other important
entropic-setting involves semiclassical physics and one employs
there the semiclassical, phase-space field Wehrl entropy (FWE)
\cite{wehrl}. In turn, the FWE has been successfully applied in
descriptions of different properties of quantum optical fields,
such as phase-space uncertainty \cite{mira1}, decoherence
\cite{deco,{orl}}, etc., a theme  that  will be the focus of our
concern in this work.  Thus, we are led to  the concept of Wehrl
phase distribution (WPD), that has been extensively developed and
shown to be a successful indicator of both noise (phase-space
uncertainty) and phase randomization \cite{mira3}. Furthermore,
the FWE has been fruitfully applied to dynamical systems too. In
this respect it must mention that the FWE-time evolution in the
case of
the Jaynes-Cummings model has been thoroughly investigated in \cite%
{orl,obad,abotalb08}. The FWE i) turns out to be more apt for
distinguishing amongst states than the NE \cite{mira3} and ii) is
known to yield helpful information on atomic inversion processes.
 Indeed,  FWE-studies of a single-trapped ion interacting with a laser
field with different configurations of the laser field have been
considered in \cite{abotalb09}. We also know now that both (a) the
fluctuations of the laser phase and (b) the initial-state setting
play important roles concerning the evolution of quantifiers like
the Husimi $Q-$function, Wehrl's entropy and Wehrl's phase
distribution\cite{abotalb09}. \vskip 3mm  {\noindent} A rather
different functional of the probability distribution function
(PDF), called Fisher's information (FI) \cite{R2} will also be
invoked here.  FI was originally introduced by Fisher \cite{R2} as
a measure of ``intrinsic accuracy" in statistical estimation
theory. We will concern ourselves in this communication with the
FI-version constructed with the
semi-classical Husimi PDF \cite{pp1,pp2,pp3}. It has been shown in \textrm{%
\cite{R12}} that FI can be used for evaluating the accuracy limits
of a quantum measurement because it provides one with meaningful
error estimates, even in the case of highly nonclassical regimes.
This is due to the fact that variances are used to quantify the
error in quantum measurements (variances and FI are intimately
linked via the Cramer-Rao bound \cite{R2}). The relation between
the so-called atomic Fisher information (AFI) and different
entanglement measures such as von Neumann's, linear, and atomic
Wehrl's entropy has been investigated in \cite{OAP10}, whose
authors found that the entanglement of a two-level atom can be
measured by using it. Also, FI is used to measure the correlation
between the quantized field and a Kerr
medium \cite{sijqi09}. A still new application for FI is found in \cite%
{OSP11}: it can be employed as an information quantifier for the
description of the weak field versus strong field dynamics in the
case of a trapped ion in a laser field. Ref. \cite{OSP11} compared
FI, as an information quantifier, with von Neumann's and Wehrl's
entropies, and provided some analytical FI-results. \vskip 3mm
{\noindent} In the present contribution, that utilizes all the
above quantifiers again, our main interest lies in a {\it
different direction}. We wish to investigate the FI-temporal
evolution for a single-qubit system in the presence of an
intensity dependent field. Now, given the pertinent PDF we can
easily build up its marginals and, in turn, a Fisher measure for
each of the resulting PDFs. We ask then: can all these FIs be used
as a quantifiers of the classical correlations and dynamical
properties of the system ar hand? Additionally, we will focus
attention on the effects of i) the laser phase and ii) the initial
state setting on the evolution of both Wehrl's entropy and
Fisher's information. Why does this matter? Because these two
measures describe (a) interesting semiclassical physics and (b)
both classical correlations and also quantum entanglement
\cite{batle,batle1,batle2}.

{\noindent} The paper is organized as follows: Section 2 deals with
preliminary matters: the basic model of a single-qubit in the presence of an
intensity dependent field together with the Wehrl entropy fundamentals, one
the one hand, and the different FIs used here on the other one. Section 4 is
devoted to the discussion of our numerical results and some conclusions are
drawn in Section 5.

\section{Preliminary materials}

\subsection{The model}

{\noindent } We give below the main details for the treatment of a two-level
atom interacting with a single-mode of the cavity field \cite{FVR98}. The
Hamiltonian, in the rotating wave approximation, can be written as \cite%
{FVR98}
\begin{equation}
\hat{H}=\omega _{F}\hat{a}^{\dagger }\hat{a}+\frac{\omega _{A}}{2}(|e\rangle
\langle e|-|g\rangle \langle g|)+\lambda f\left( \hat{a}^{\dagger }\hat{a}%
\right) (\hat{a}^{\dagger }|1\rangle \langle 0|+\hat{a}|0\rangle \langle 1|),
\label{1}
\end{equation}%
where $\omega _{F}$\ is the field frequency, $\omega _{A}$ the
transition frequency between the upper $|0\rangle $ and lower
state $|1\rangle $ of the atom, and $\lambda $\ the effective
coupling constant. The field creation
(annihilation) operator is $\hat{a}^{\dagger }\left( \hat{a}\right) $ while $%
f(\hat{a}^{\dagger }\hat{a})$ represents the intensity dependent function of
the cavity field mode. Restricting ourselves to the functional form $f(\hat{n%
})=\sqrt{\hat{a}^{\dagger }\hat{a}}$, the interaction Hamiltonian reads
\begin{equation}
\hat{H}_{I}=\lambda (\hat{\psi}^{\dagger }|1\rangle \langle 0|+|0\rangle
\langle 1|\hat{\psi}),  \label{hi}
\end{equation}%
where $\hat{\psi}=\hat{a}\sqrt{\hat{a}^{\dagger }\hat{a}}$ and $\hat{\psi}%
^{\dagger }=\sqrt{\hat{a}^{\dagger }\hat{a}}\hat{a}^{\dagger }$. The time
evolution operator for the effective Hamiltonian (2) becomes
\begin{equation}
U^{^{\dagger }}(t)=\exp [-i\hat{H}_{I}t]=\left[
\begin{array}{cc}
\cos \left( T\sqrt{\hat{\psi}\hat{\psi}^{\dagger }}\right) & i\dfrac{\sin
\left( T\sqrt{\hat{\psi}\hat{\psi}^{\dagger }}\right) }{\sqrt{\hat{\psi}\hat{%
\psi}^{\dagger }}}\hat{\psi} \\
i\dfrac{\sin \left( T\sqrt{\hat{\psi}^{\dagger }\hat{\psi}}\right) }{\sqrt{%
\hat{\psi}^{\dagger }\hat{\psi}}}\hat{\psi}^{\dagger } & \cos \left( T\sqrt{%
\hat{\psi}^{\dagger }\hat{\psi}}\right)%
\end{array}%
\right].
\end{equation}
{\noindent } In this last relation  $T=\lambda t$ \ is\ the scaled
time. The time units are given by the inverse of the coupling
constant $\lambda $. We assume (I) that the initial state of the
system is the product $\rho^{AF}\left( 0\right) =\rho ^{A}\left(
0\right) \otimes \rho^{F}\left( 0\right) $, with our qubit
assigned initially to the upper state, i.e., $\rho ^{A}\left(
0\right) =|0\rangle \left\langle 0\right\vert $, while (II) the
field's initial state is a coherent-one $\rho^{F}(0)=|\alpha
\rangle \langle \alpha |=\sum\limits_{n,m=0}^{\infty }C_{n}\left(
\alpha \right) C_{m}^{\ast }\left( \alpha \right) |n\rangle
\left\langle m\right\vert, $ with \be C_{n}\left( \alpha \right)
=\alpha ^{n}\sqrt{\dfrac{e^{-|\alpha |^{2}}}{n!}}. \nonumber\ee
The Husimi $Q-$function $Q_{F}$ of the field-mode, in terms of the
diagonal elements of the density operator in the coherent-state
basis, is

\begin{equation}
Q_{F}\left( \beta ,t\right) =\dfrac{1}{\pi }\text{Tr}_{A}\left\langle \beta
\left\vert U(t)\rho ^{AF}(0)U^{\dagger }(t)\right\vert \beta \right\rangle
\text{,}  \label{w1}
\end{equation}%
\ where Tr$_{A}$ means that we trace over the atomic variables.
\vskip 3mm

\noindent Next, we turn our attention to the semiclassical
Wehrl-entropy \cite{wehrl} that describes the time evolution of a
quantum system in phase-space. This entropy, introduced as the
classical entropy of a quantum
state, yields meaningful insights into the dynamics of the system \cite%
{wehrl} and is defined as the coherent-state representation of the density
matrix \cite{wehrl, orl} via

\begin{equation}
S_{W}\left( t\right) =-\int Q_{F}\left( \beta ,t\right) \ln Q_{F}\left(
\beta ,t\right) d^{2}\beta ,
\end{equation}%
where $d^{2}\beta =\left\vert \beta \right\vert d\left\vert \beta
\right\vert d\Theta $ . We can specialize things by recourse to the Wehrl
phase distribution (Wehrl PD), defined to be the phase density of the Wehrl
entropy \cite{mira3,abotalb09}, i.e.,

\begin{equation}
S_{\Theta }(t)=-\int Q_{F}\left( \beta ,t\right) \ln Q_{F}\left( \beta
,t\right) \left\vert \beta \right\vert d\left\vert \beta \right\vert
\end{equation}%
where $\Theta =\arg \left( \beta \right) $.

\subsection{Fisher Information}

{\noindent } The Fisher information measure (FIM) for any PDF
$f(\mathbf{x})$ can be cast in the fashion \cite{roybook,5}
\begin{equation}
I=\int d\mathbf{x\ }f(\mathbf{x})\left\{ \frac{\partial \ln f(\mathbf{x})}{%
\partial \mathbf{x}}\right\} ^{2},
\end{equation}%
and is encountered in many physical applications (see, for
instance, \cite{KSC10}-\cite{FlaviPRE}, and references therein).
The FIM associated to Husimi distributions $Q_{F}(X_{1},X_{2},t)$
is defined as \cite{PP04}
\begin{eqnarray}
I_{F}(t) &=&\dfrac{1}{2\pi \hbar }\int_{-\infty }^{\infty }\int_{-\infty
}^{\infty }\mathbf{\ }Q_{F}(X_{1},X_{2},t)\Gamma (X_{1},X_{2},t)dX_{1}dX_{2},
\notag \\
&=&\int Q_{F}\left( \beta ,t\right) \Gamma \left( \beta ,t\right) d^{2}\beta
,  \label{FF1}
\end{eqnarray}%
where $\Gamma (X_{1},X_{2},t)$ and $\Gamma \left( \beta ,t\right)
$ can be written in terms of the phase space parameters, yielding

\begin{equation}
\Gamma (X_{1},X_{2},t)=\sum\limits_{j=1}^{2}\left( \sigma _{X_{j}}(t)\frac{%
\partial \ln (Q_{F}\left( X_{1},X_{2},t\right) )}{\partial _{X_{j}}}\right)
^{2},
\end{equation}

\begin{equation}
\Gamma \left( \beta ,t\right) =\underset{j=1}{\overset{2}{\sum }}\sigma
_{X_{j}}^{2}(t)\underset{k=1}{\overset{2}{\sum }}\left( \frac{\cos \left(
\Theta +\dfrac{\pi }{j}-\dfrac{\pi }{k}\right) }{\left[ k-2+\beta \left(
k-1\right) \right] }\frac{\partial \ln (Q\left( \beta ,t\right) )}{\partial %
\left[ \beta \left( k-2\right) +\Theta \left( k-1\right) \right] }\right)
^{2},  \label{gamb}
\end{equation}%
with  \be \label{variance} \sigma _{X_{j}}(t)=\sqrt{\left\langle
X_{j}(t)^{2}\right\rangle -\left( \left\langle
X_{j}(t)\right\rangle \right) ^{2}}, \ee and
\begin{equation*}
\left\langle X_{j}\right\rangle =\int_{-\infty }^{\infty }\int_{-\infty
}^{\infty }\mathbf{\ }X_{j}Q(X_{1},X_{2},t)dX_{1}dX_{2}.
\end{equation*}%
We also consider, as a dynamical measure, the quantity

\begin{equation}
I_{\Theta }(t)=-\int_{0}^{\infty }Q_{H}\left( \beta ,t\right)
\Gamma \left( \beta ,t\right) \left\vert \beta \right\vert
d\left\vert \beta \right\vert . \label{fased} \end{equation}

\bigskip {\noindent } It is worth noting that the definition (\ref{FF1}) is
given in analogy to that of the field Wehrl entropy (the special case $t=0$
being $I_{F}(0)=2$) so that the corresponding Fisher's phase distribution
can be cast, in terms of the error function $\mathrm{erf}(x)=\frac{2}{\sqrt{%
\pi }}\int_{0}^{x}e^{-y^{2}}dy$, as
\begin{eqnarray}
I_{\Theta }(0) &=&\frac{1}{2\pi }\exp \left( x^{2}-\alpha ^{2}\right)  \notag
\\
&&\times {\{x\sqrt{\pi }[1+\mathrm{{erf}\left( x\right) }]f_{1}}+\exp \left(
-x^{2}\right) f_{2}{\}},  \label{fw66}
\end{eqnarray}%
where $x=\alpha \cos (\Theta )$ and
\begin{equation}
f_{j}=\alpha ^{2}-x^{2}+\frac{j}{2},\;\;j=1,2\text{\ }.
\end{equation}%
In correlations terms the bipartite system becomes uncorrelated whenever $%
I_{F}(0)\simeq 2$, this value representing the lower bound for $I_{F}$. One
has

\bigskip
\begin{equation}
I_{F}(0)=S_{W}(0)+1-\ln \pi ,  \label{con1}
\end{equation}%
and
\begin{equation}
I_{\Theta }(0)=S_{\Theta }(0)-\frac{\ln \pi }{2\pi }\exp \left( -\alpha
^{2}\right) {\biggl\{}1+{x\sqrt{\pi }\left( 1+\mathrm{{erf}\left( x\right) }%
\right) }\exp (x^{2}){\biggr\}}.  \label{con2}
\end{equation}%
Equations (\ref{con1}) - (\ref{con2}) establish the connection between
Fisher's information measure (FIM) and Wehrl's entropy at $T=$ $0$. Notice
that at this time $I_{F}-I_{\Theta }=1-\ln \pi =$ constant, which is a
counterintuitive result, since one expects their sum to be approximately
constant \cite{FlaviPRE}. This curious result is due to the periodicity of
the evolution.

\section{Results}

\begin{figure}[tbp]
\centering\includegraphics[width=0.8\textwidth]{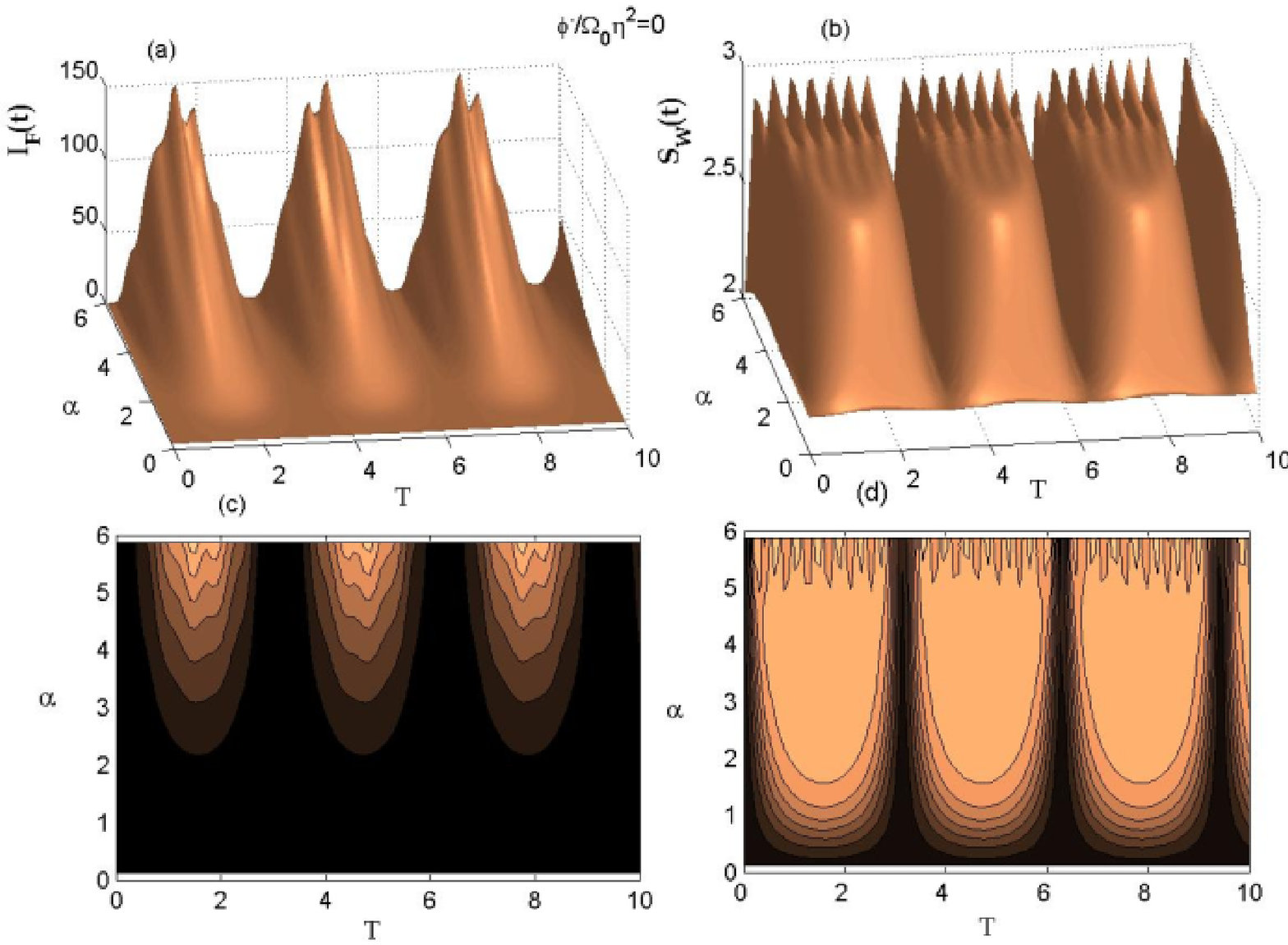}
\caption{The volution of (a,c) Fisher' information $I_{F}(t)$,
(b,d) Wehrl's entropy $S_{W}(t)$, versus the scaled time
$T=\protect\lambda t$ and the root of the mean photon number
$\protect\alpha =\protect\sqrt{\bar{n}}$.} \label{figp1}
\end{figure}

\noindent We start now the presentation of our numerical results.
We will see that the coherent state parameter $\alpha $,
representing the square root of the mean-photon number, greatly
influences the dynamics, as can be clearly appreciated in Figs.
\ref{figp1} that depict, respectively, $I_{F}$ and $S_{W}$ as a
function of $T$ and $\alpha $ [(a) and (b)] together with their
projections on the $\alpha -T-$plane [(c) and (d)] ($T$ is a
\textquotedblleft scaled" time). Both the inherent periodicity of
the dynamics and the long living correlation between the single
qubit and the coherent field are clearly visible. They increase as
the phonon-number grows. Both quantifiers exhibit the periodicity
of the system.

\begin{figure}[tbp]
\centering\includegraphics[width=0.8\textwidth]{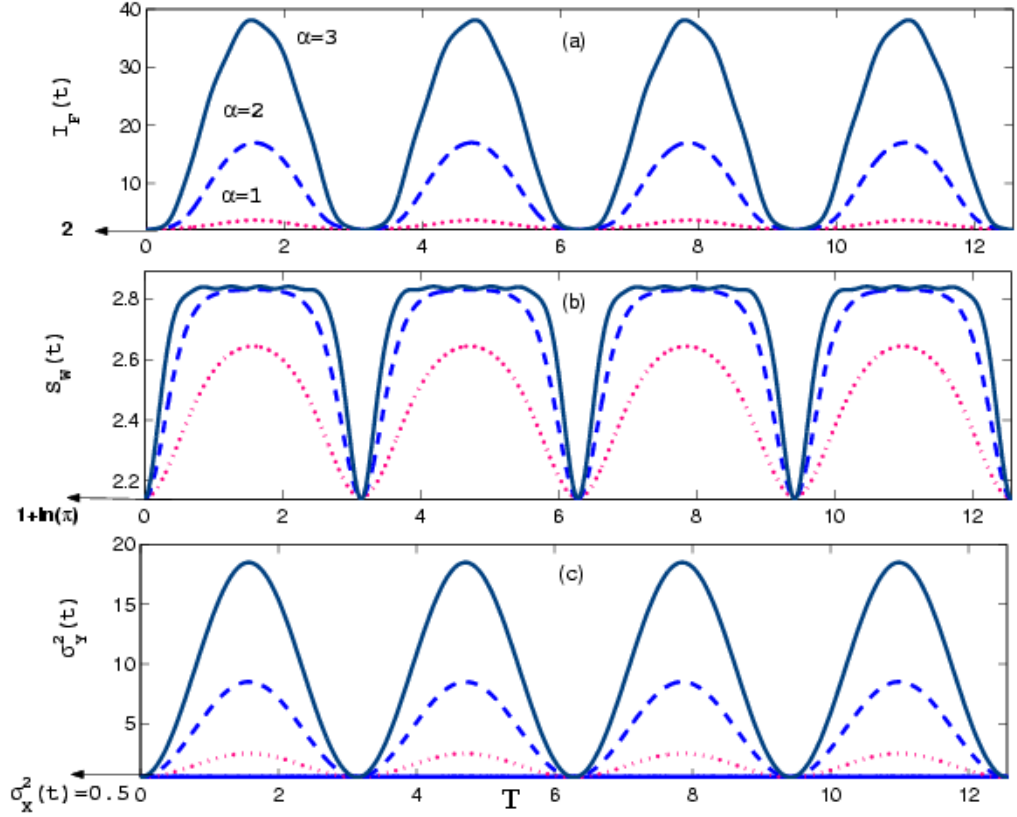}
\caption{Time volution of (a) Fisher' information $I_{F}(t)$, (b)
Wehrl's entropy $S_{W}(t)$, and (c) variances $\protect\sigma
_{X_2}^{2}(t)$. We set $Y\equiv X_2$. We also plot
curves for different values of the square root of the mean photon number $%
\protect\alpha ,$ namely, dotted curve for $\protect\alpha =1$,
dashed curve for $\protect\alpha =2$, and solid curve for
$\protect\alpha =3$.} \label{Khg1}
\end{figure}

\nd In figure \ref{Khg1} we plot the FI- and Wehrl- time
evolutions together with the associated  $Y-$variance. For
typographical simplicity, we set in the graphs
$Y\equiv X_2.$ One chooses three values of the $\alpha-$%
parameter, namely, $=1,2,3$, respectively. In order to ensure good
accuracy, the behavior of the Fisher information $I_{F}(t)$ has
been determined using an appropriate scale so as to meaningfully
compare it to Wehrl's entropy. FI's behavior is clearly dominated
by the variance component $\sigma _{X_2}^{2}(t)$ [Cf. Eq.
\ref{variance}].

\begin{figure}[tbp]
\centering\includegraphics[width=0.8\textwidth]{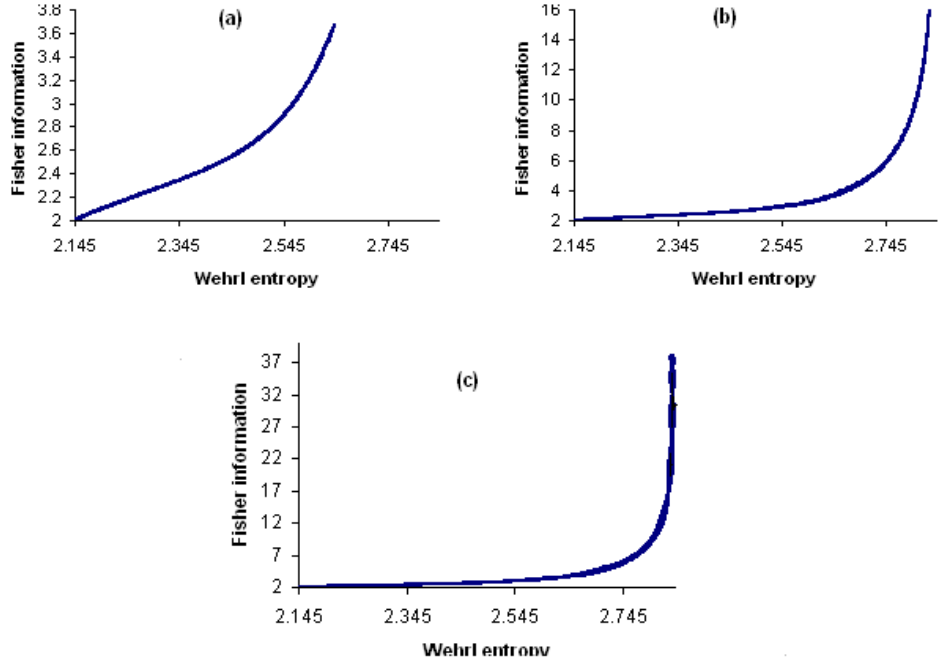}
\caption{Illustration of the $I_{F}$ vs. $S_{W}$ \ behavior for
parameters values for different values of \ the square root of
$\protect\alpha $, where
figure (a) $\protect\alpha =1$, figure (b) $\protect\alpha =2$, and $\protect%
\alpha =3$ for figure (c).} \label{Khg6}
\end{figure}

{\noindent }Fig. \ref{Khg6} illustrates the $I_{F}$ vs. $S_{W}$
behavior, which is surprising indeed. The second is a global
measure, while the former is a local one \cite{ppglobal}. One
expects them to behave in opposite fashion \cite{ppglobal}. Here
though, a monotonous comportment is apparent, for the first time
as far as we know. Rapid fluctuations, combined with strong
delocalization are thus generating a bizarre scenario. There is
more to be said, though. Wherl's entropy was invented as a measure
of delocalization in phase-space. For the minimum possible mean
photon number (corresponding to $\alpha=1$), quantum indeterminacy
is maximal and both quantifiers grow together. We may understand
this behavior if we realize that, as indeterminacy augments (and
so does $S_W$) fluctuations rise, forcing $I_F$ to increase as
well. Enters now a bit of rather interesting physics. The route to
classicality is paved by i) a growing mean photon-number and ii) a
stabilization of Wehrl's entropy since delocalization stops
augmenting. This scenario begins to insinuate itself at $\alpha=2$
and becomes fully installed already at $\alpha=3$! Thus, $S_W$
cannot grow, but nothing impedes $I_F$ to increase. What we are
really  watching  in Fig. \ref{Khg6}-(c) is the emergence of the
classical limit.

\begin{figure}[tbp]
\centering\includegraphics[width=0.8\textwidth]{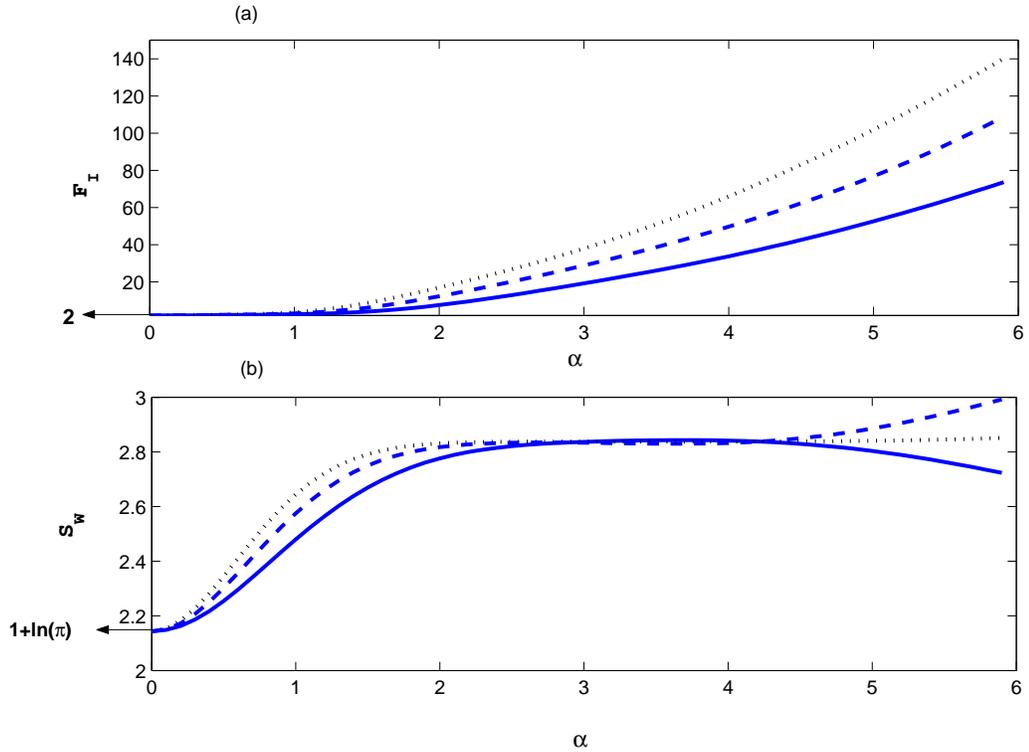}
\caption{We depict the FIM-Wehrl behaviors versus the square root
of the mean photon number. } \label{Khg4}
\end{figure}

{\noindent} The above considerations receive a boost via Fig.
\ref{Khg4}, that  depicts the FIM-Wehrl behavior versus the mean
phonon number. In a sense, things return to ``normality", as the
quantifiers behave differently here, as expected. $I_{F}$ always
grows with $\alpha $ entailing that, as one intuitively
understands, errors diminish as particle-number grows. Wherl's
measure has a peak at about $\alpha =3$. This can also be
understood. $S_{W}$ measures our ignorance about localization in
phase-space, which, as it should, diminishes as $\alpha $ grows.

\subsection{Cramer-Rao bound}

\begin{figure}[tbp]
\centering\includegraphics[width=0.8\textwidth]{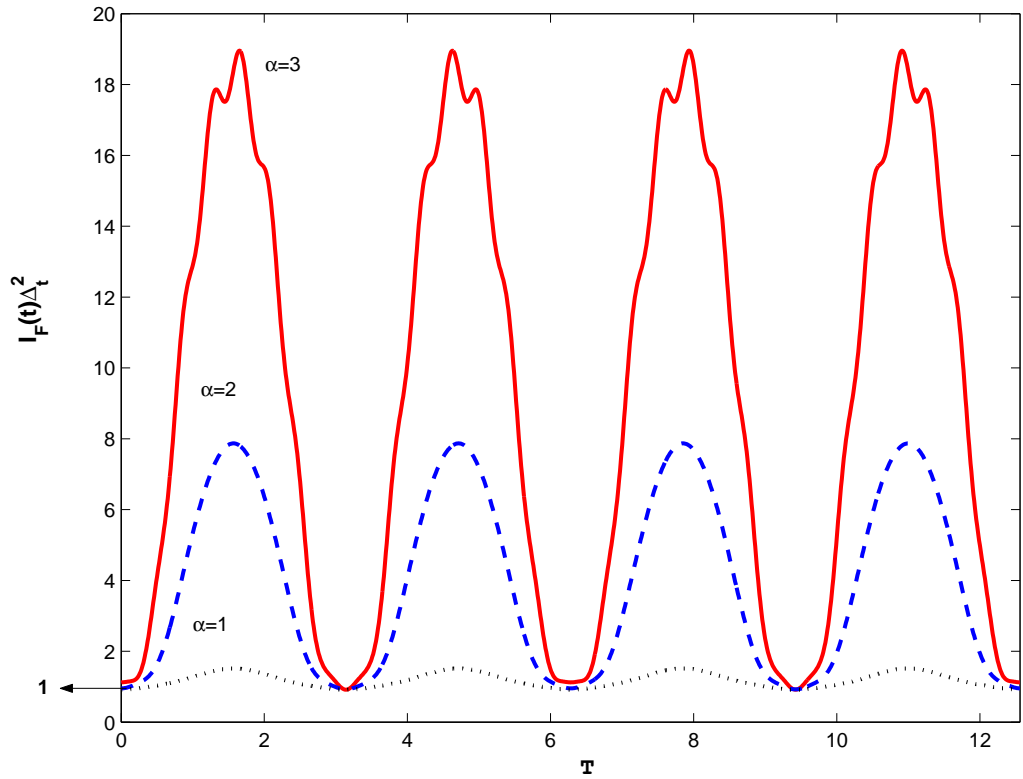}
\caption{The Cramer-Rao product $I_{F}(t)\Delta _{t}^{2}$ as a
function of the scaled time $T$ \ for different values of \
$\protect\alpha $ where the dotted curve $\protect\alpha =1$,
dashed curve $\protect\alpha =2$ and the solid curve
$\protect\alpha =3$.} \label{CRfig5}
\end{figure}

{\noindent} The ``true" FI-informational content is conveyed by the
Cramer-Rao inequality (CR). Indeed, this is its most important property. We
recapitulate in one-dimension, for simplicity's sake. If the classical
Fisher information associated with translations of a one-dimensional
observable $x$ with corresponding probability density $f(x)$ is~\cite%
{roybook,Hall}

\begin{equation}
I_x=\int \mathrm{d}x\,f(x)\,\left(\frac{\partial \ln f(x)}{\partial x}%
\right)^2,
\end{equation}
then it obeys the above referred inequality, namely

\begin{equation}
(\Delta x)^2\geq I_x^{-1}  \label{xcramer1}
\end{equation}
involving the variance of the stochastic variable $x$~\cite{Hall}

\begin{equation}
(\Delta x)^{2}=\langle x^{2}\rangle -\langle x\rangle ^{2}=\int \mathrm{d}%
x\,f(x)\,x^{2}-\left( \int \mathrm{d}x\,\rho (x)\,x\right) ^{2}.
\end{equation}%
We remark that the derivative operator significantly influences the
contribution of minute local $f-$variations to FI's value, so that the
quantifier is called a \textquotedblleft local" one. Note that Wehrl's
entropy decreases with skewed distributions, while Fisher's information
increases in such a case. Local sensitivity is useful in scenarios whose
description necessitates appeal to a notion of \textquotedblleft order". {%
\noindent} For our present purposes we deal with a time-dependent CR, that,
in self-explanatory notation reads

\begin{equation*}
I_{F}(t)\Delta _{t}^{2},
\end{equation*}
where

\begin{equation*}
\left.
\begin{array}{c}
\Delta _{t}^{2}=\left\langle \beta ^{2}\right\rangle -\left\langle \beta
\right\rangle ^{2} \\
\left\langle \beta ^{s}\right\rangle =\int_{0}^{2\pi }\int_{0}^{\infty
}\left\vert \beta \right\vert ^{s}Q_{F}\left( \beta ,t\right) \left\vert
\beta \right\vert d\left\vert \beta \right\vert d\Theta ;\text{ \ \ \ \ \ \
\ \ \ \ }s=1,2,%
\end{array}%
\right.
\end{equation*}%
and we display in Fig. \ref{CRfig5} some results. We see that the
CR product oscillates with time and rapidly decreases as $\alpha $
grows. At $\alpha=3$ the product almost reaches its lower bound of
unity, reconfirming the fact that at that $\alpha-$value
classicality has been reached.

\section{Conclusions}

{\noindent } We have here considered, from an information-theory viewpoint,
the dynamics of a single-qubit system. Our information-quantifiers were the
Wehrl entropy, a phase-space delocalization measure, and Fisher's
information. These two quantities aptly illustrate on the complicated
dynamics at hand. The main characteristics of the problem are governed by
the mean phonon number and the intensity dependent field.

{\noindent } We have performed an extensive numerical analysis, illustrated
via a variety of graphs. Periodicity is a main feature, specially as the
mean-phonon number grows. Long living correlations between the qubit system
and and coherent field are clearly appreciated. The monotonous growing of
the Fisher measure as the Wehrl entropy grows is a counterintuitive feature
that we have detected. This is a surprising facet because it is well known
that whenever Fisher grows, Shannon or Wehrl entropies decrease \cite%
{roybook,FlaviPRE}. Notice that FI measures gradient content \cite%
{roybook,FlaviPRE} while Wehrl's measure is a delocalization-indicator \cite%
{FlaviPRE}. However, the physics of the phenomenon can be
understood, as explained in the preceding Section. Rather
unexpectedly, we get illuminating insights into the emergence of
classicality,  in line with very recent findings \cite{nowadays}.

\newpage

\end{document}